\documentclass{webofc}

\usepackage{listings}
\usepackage[draft]{minted}
\usepackage[caption=false]{subfig}
\usepackage{hyperref}
\usepackage{graphicx}
\usepackage{tikz}

\usepackage{minted}

\usepackage[varg]{txfonts}   

\newminted{cpp}{gobble=2, numbersep=3pt, escapeinside=||, xleftmargin=8mm, framesep=5mm, frame=leftline, baselinestretch=1, fontsize=\scriptsize, linenos=false}
\newmintinline{cpp}{}

\usetikzlibrary{shadows, arrows, fit, positioning, matrix, shapes}
\usetikzlibrary{decorations.markings}

\definecolor{my_yellow}{RGB}{255, 253, 217}
\definecolor{my_orange}{RGB}{255, 127, 0}
\definecolor{my_lightblue}{RGB}{105, 186, 249}
\definecolor{my_purple}{RGB}{150, 154, 219}
\definecolor{my_green}{RGB}{90, 194, 160}

\tikzset {
  bigbox/.style = {draw, thick, fill=gray!10, rounded corners, rectangle},
  box/.style = {draw, thick, minimum height=0.8cm, minimum width=1.5cm, rounded corners, rectangle, fill=white, anchor=south},
  model/.style = {draw, thick, fill=white, text centered, minimum height=3em, minimum width=4em, rounded corners, drop shadow},
  user/.style = {draw, thick, ellipse, fill=white, text centered, minimum height=3em, minimum width=5em, drop shadow},
  line/.style = {->, thick, color=black, shorten <=2pt, shorten >=2pt, >=stealth'},
  dashed/.style = {->, dash pattern=on 3pt off 3pt, color=gray, shorten <=2pt, shorten >=2pt, >=stealth'},
  plain/.style = {minimum width=1em},
  arcnode/.style 2 args={
    decoration={
                 raise=#1,             
                 markings,   
                 mark=at position 0.5 with {\node[inner sep=0] {#2};}
            },
            postaction={decorate}
    }
}

\pgfdeclarelayer{background}
\pgfdeclarelayer{foreground}
\pgfsetlayers{background,main,foreground}

\begin{document}
\title{Extending ROOT through Modules}
%
%

\author{\firstname{Oksana} \lastname{Shadura}\inst{1}\fnsep\thanks{\email{oksana.shadura@cern.ch}} \and
        \firstname{Brian Paul} \lastname{Bockelman}\inst{1}\fnsep\thanks{\email{bbockelm@cse.unl.edu}} \and
        \firstname{Vassil} \lastname{Vassilev}\inst{2}\fnsep\thanks{\email{vvasilev@cern.ch}}
}

\institute{University of Nebraska Lincoln, 1400 R St, Lincoln, NE 68588, United States
\and
           Princeton University, Princeton, New Jersey 08544, United States
          }

\abstract{%
The ROOT software framework is foundational for the HEP ecosystem, providing capabilities such as IO, a C++ interpreter, GUI, and math libraries. It uses object-oriented concepts and build-time components to layer between them. We believe additional layering formalisms will benefit ROOT and its users.

We present the modularization strategy for ROOT which aims to formalize the description of existing source components, making available the dependencies and other metadata externally from the build system, and allow post-install additions of functionality in the runtime environment. components can then be grouped into packages, installable from external repositories to deliver post-install step of missing packages. This provides a mechanism for the wider software ecosystem to interact with a minimalistic install. Reducing intra-component dependencies improves maintainability and code hygiene. We believe helping maintain the smallest “base install” possible will help embedding use cases.

The modularization effort draws inspiration from the Java, Python, and Swift ecosystems. Keeping aligned with the modern C++, this strategy relies on forthcoming features such as C++ modules. We hope formalizing the component layer will provide simpler ROOT installs, improve extensibility, and decrease the complexity of embedding in other ecosystems.
}
\maketitle
\section{Introduction}

One clear advantage of object-oriented systems is the ability to have abstraction layers provide separation between the user of an object and the implementer. Ultimately, this helps projects scale to very large code bases. Another technique for scaling projects is using modularization: grouping together significant functionality into distinct units with clear points of interaction.  While ROOT has a strong history in object-oriented programming, it doesn’t have a strong concept of components.  Thus, we propose to add four concepts to the ROOT ecosystem:
\begin{enumerate}
\item Component: A set of interdependent classes implementing coherent functionality and providing well-defined APIs.
\item Library: a component or set of components  which makes sense to be together and that can be used in a program or another library.
\item Package: A distinct, self-describing resource (file, URL) that provide one or more components.
\item Package database: A record of all packages currently available in a ROOT installation.
\item Package manager: An actor that can locate and install packages into a ROOT installation from a package reference, along with their transitive dependencies.
\end{enumerate}
There exist other legacy large object-oriented software systems  which, similar to ROOT, consisting of a large number of interdependent  and loosely-coupled classes, mostly organized as set of libraries and build targets.  Classes are often the lowest level of granularity to serve as a unit of software modularization. In ecosystems such as Java, Python and C++, a further package structure can allow software developers organize their programs into components. A good organization of classes into identifiable and collaborating packages eases the understanding, maintenance of software. To improve the quality of software modularization, assessing the package organization is required.

Since we address large software systems as ROOT framework, consisting of a very large number of classes and packages, we consider that packages are the units of software modularization. 

Packages and components are not synonymous concepts.  A component is a set of functions, types, classes and etc., defined in a common namespace. We can define component also as a set of related functionality that exposes a well-defined API. A library is a component or set of components  which makes sense to be together and that can be used in a program or another library. A package is a unit of distribution that can contain a library or an executable or both. Considering a library is a set of components, a package bundles a set of components with related metadata (such as name, version and dependencies) which can be distributed as a standalone unit.

\section{Motivation}

Modularization defines a way of grouping of functionality from a software product. It outlines groups in form of components which identify a particular piece of functionality to solve a set of problems. In general, modularization helps reducing management, coordination and development costs. We aim to define a set of mechanisms that enables a modular version of ROOT, centered around C++ modules, working as package-centric ecosystem.

Library dependencies alone result in an imposingly complex relationship diagram. By introducing a component layer, we would provide better boundaries between components, allowing ROOT to scale as a project. For example, the level of expertise for the contributor needs can be more localized. It could means for project, that ROOT by itself could evolve in new phase, and can potentially interact with many more packages and while turn itself into even more useful toolkit. 

By making the boundaries and relationships more explicit through components, we can better define and implement a “minimal ROOT,” increasing the chances its functionality can be embedded in other contexts.  This enables ROOT users to interact with the wider data science ecosystem.

Packages and package management provide a mechanism for ROOT users to socialize and and reuse projects built in the context of ROOT, it helps to make ROOT more flexible and open for new customers.  This allows ROOT to continue to serve as a community nexus.

In particular, this provides the ROOT team with an improved mechanism to say “no” to new components within the ROOT source itself as users can simply share their packages among each other or in a common store such as github.

\section{ROOT components and packages}

ROOT code is organized into a set of logical sub-directories, except of Core library, that has more complex structure of subfolders. In this case, each of directories can be roughly thought of as a component. 

A component is a single unit of code distribution or set of classes for a framework or an application that is built and ready to be shipped. It can be imported by another C++ module \cite{cxxmodules} or other component with a hook, such as  include in case of headers or an import keyword. A program may have all of its code in a single component (or inside of more components), or it may import other components as dependencies.

This model requires a few definitions:
\begin{enumerate}
\item Define components as a ROOT installation “unit". It could be also defined as a component or package with single component inside which is distributed and maintained through 'standard' ROOT (i.e. lives in root.git). \cite{rootgit}

For example, we could define a Minuit ROOT component as a subsystem for numerical minimization,used for a functionality of search of extreme points and is intended to be included as a part of a finished, packaged, and labeled package for providing a one-dimensional and multidimensional minimization for ROOT ecosystem.

As an other example, Figure 1 shows possibility to combine any type custom package based on user required components, like for example hypothetical ROOT-ML package that consists of TMVA and RDataframe components with its dependencies.

\item Every facility is provided by exactly one component.
\end{enumerate}

Taking in account ongoing work on C++ modules \cite{rootcxxmodules} in ROOT, we aligned the work together with ROOT modularization efforts. On one hand, C++ modules provide an simple way to use software libraries that provides better compile-time scalability and eliminates problems to access the API of a library. They improve encapsulation and outline a well-defined relationship between public and private part of the code split into an implementation and interface. On the other hand, modules have a weak side in versioning and binary distribution of modules among others problems.


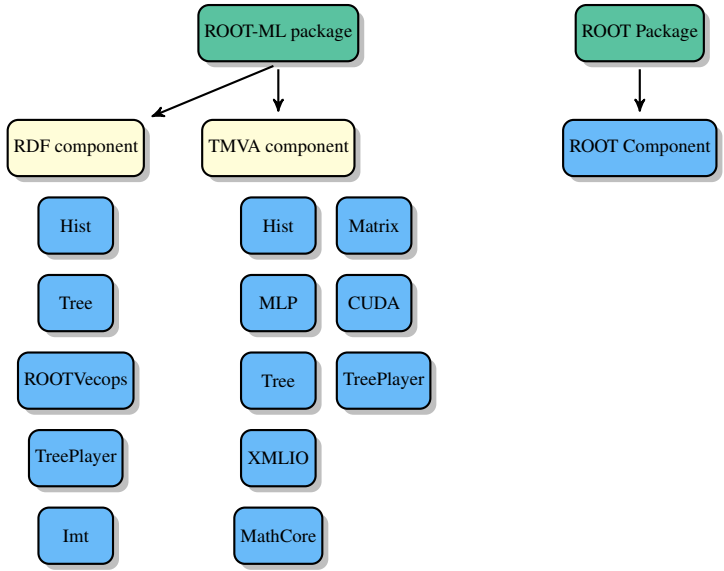
\begin{figure}[!h]
  \centering
    \begin{tikzpicture}[outer sep=0.05cm, node distance=0.8cm, scale=0.7, transform shape]
    \node[model, fill=my_green, name=rootml] (rootml) {ROOT-ML package};   
    \node[model, fill=my_green, name=pkg, right=4cm of rootml] (pkg) {ROOT Package};
    \node[model, fill=my_lightblue, name=component, below=1cm of pkg] (component) {ROOT Component};
    \node[model, fill=my_yellow, name=tmva, below=1cm of rootml] (tmva) {TMVA component};
    \node[model, fill=my_yellow, name=rdf,  left=1cm of tmva] (rdf) {RDF component};
    \node[model, fill=my_lightblue, name=hist, below=0.3cm of tmva] (hist) {Hist};
    \node[model, fill=my_lightblue, name=mlp, below=0.3cm of hist] (mlp) {MLP};
    \node[model, fill=my_lightblue, name=tree, below=0.3cm of mlp] (tree) {Tree};
    \node[model, fill=my_lightblue, name=xmlio, below=0.3cm of tree] (xmlio) {XMLIO};
    \node[model, fill=my_lightblue, name=mathcore, below=0.3cm of xmlio] (mathcore) {MathCore};
    \node[model, fill=my_lightblue, name=matrix, right=0.3cm of hist] (matrix) {Matrix};
    \node[model, fill=my_lightblue, name=cuda, right=0.3cm of mlp] (cuda) {CUDA};
    \node[model, fill=my_lightblue, name=tp, right=0.3cm of tree] (tp) {TreePlayer};
    
    \node[model, fill=my_lightblue, name=hist1, below=0.3cm of rdf] (hist1) {Hist};
    \node[model, fill=my_lightblue, name=tree, below=0.3cm of hist1] (tree) {Tree};
    \node[model, fill=my_lightblue, name=vecops, below=0.3cm of tree] (vecops) {ROOTVecops};
    \node[model, fill=my_lightblue, name=xmlio, below=0.3cm of vecops] (tpl) {TreePlayer};
    \node[model, fill=my_lightblue, name=mathcore, below=0.3cm of tpl] (imt) {Imt};
    
    \draw[line, ->] (pkg.south) -- (component);
    \draw[line, ->] (rootml.south) -- (tmva);
    \draw[line, ->] (rootml.south) -- (rdf);

  \end{tikzpicture}
  \caption{Example of ROOT component and package.}
  \label{fig:package}
\end{figure}

ROOT package is defined as a grouping of software for data analysis and associated resources, intended for it distribution (extension or upgrade of  ROOT functionality). 
The definition of package assumes a contract for code organization in order to simplify the build and deploy steps. The contract defines a manifest file and particular organization of each component.

Manifest file is a file which describes the content of a package. It has self-describing and easy to process by machines format. The manifest file contains information about how the contents should be built, deployed and versioned. Example of manifest file is shown in Listing \ref{manifest}).

\begin{listing}[h]
\noindent
\begin{minipage}[h]{.7\textwidth}
\begin{cppcode*}{}
  package:
    name: "ROOTMath"
    targets:
      target:
      name: "MathCore MathMore mathcore-tests mathmore-tests"
    products:
      package:
      name: ROOTMath
      targets: MathCore MathMore VecCore Imt gsl
    module:
      name: MathCore
      publicheaders: inc/<enumerated headers>.h
      sources: src/<enumerated source files>.cxx
      targets: MathCore
      dependencies: VecCore Imt
      tests: mathcore-tests
    module:
      name: MathMore
      publicheaders: inc/<enumerated headers>.h
      sources: src/<enumerated source files>.cxx
      targets: MathMore
      dependencies: gsl MathCore
      tests: mathmore-tests
    module:
      name: VecCore
      packageurl: "https://github.com/root-project/veccore/archive/v0.5.1.zip"
      targets: VecCore
      tag: 0.5.1
    module:
      name: gsl
      packageurl: "https://github.com/ampl/gsl/archive/v2.5.0.zip"
      targets: gsl
 \end{cppcode*}
 \end{minipage}
 \caption{Draft version of YAML manifest file for ROOTMath package.}
 \label{manifest}
\end{listing}

Package byproducts are a set of components, which could be packed as a library and/or executable, with documentation and unit tests.

One of simple case of ROOT package is ROOT Base, which include Cling, IO and Core components. This is our fundamental part from which we start to build a ROOT. Other example of ROOT package is abstract “ROOTMath” package, that could consist of multiple math related components depending on package vendor (check example in Listing \ref{manifest}).

To have a successful socialization of ROOT project via modularization, we need to agree on a format of ROOT package and define set of available components and its packages for existing monolithic ROOT framework (e.g. manifest example was inspired by Swift manifests \cite{swift} and should be written in YAML).

To provide motivation for introducing manifest files, we will to try to list its use-case scenarios:
\begin{enumerate}
\item Situation when we need to deal with ROOT subsystem user or developer (e.g. I/O developer). The manifest file is generated by the info in the build system.
\item Situation when we need to deal with third party developer (PhD student) who has some amount of files and does not know anything about build systems and in the same would like to describe in a human form what package does and what ROOT components it depends on.
\item Situation when we need to deal with experiment librarian that knows exactly what he need - writing manifest file or some other configuration to tell ROOT what packages he need for the ideal scenario. The other scenario could be to describe a pre-built package.
\item Situation when we need to deal with member of physics group that want to have particular library to be auto-build on demand or to socialize his own library.
\end{enumerate}



\section{Package manager design prototype}

\subsection{Evolving “ROOT Minimal” to “ROOT Base”}

While using CMake generation of libraries dependency graph, we can easily notice, that it is extremely hard to define “heart” part of ROOT. We need to start to work on modularization of ROOT package from the bottom to the top, trying to to build new ROOT modular system step by step.

ROOT’s build system provides a “minimal ROOT” option that suppose to deliver basic or essential functionality of ROOT, required for basic I/O and data analysis operations.

After evaluation of components built when this option is enabled, we believe that “ROOT Minimal” has migrated away from its original goal of a “core-like” ROOT; we aim to trim it down and to what we will consider the “base component” in this article. Our plan is to start with three components: Core, RIO and Cling.

ROOT Base is formed by taking these three components and their transitive dependencies using CMake-based introspection. TCling, an interpreter interface in ROOT, can help to provide an API, to contribute a set of operations with interpreter, including generation of the dictionary for the C++ classes, allowing operations with mangled names for a method of a class and etc.

Other problem is how PM will be connected to ROOT interpreter. This is where CMake falls short as it does not have any support for steps happening after build/install time PM allows bootstrapping minimal ROOT and installing packages automatically on demand. It provides a basic interpreter functionality,  which will allow to install on demand ROOT components and use them directly in the same session.

\subsection{ROOT Package Manager and "lazy install" method}

We think about ROOT package manager project/application dependency manager (PDM), since ROOT has incorporated C++ interpreter in its code source.

From definition, project/application dependency manager (PDM) is an interactive system for managing the source code dependencies of a single project in a particular language. That means specifying, retrieving, updating, arranging on disk, and removing sets of dependent source code, in such a way that collective coherency is maintained beyond the termination of any single command. 
Its output — which is precisely reproducible — is a self-contained source tree that acts as the input to a compiler or interpreter. You might think of it as “compiler, phase zero.” \cite{pdm}

A ROOT package managing system can manage the package lifetime to ensure sustainability in a transparent to the user way. The role of the ROOT package manager is to reduce coordination costs by automating the process of downloading and building all of the dependencies for a project required for users, experiments or total community.

One of main challenges is to define package granularity. The best strategy is to left this decision for the users. Packages should not contain too little and too big components because this in a way defeats the purpose of modularization. In the same time packages should not contain too many and too small components because this introduces a lot of package management overhead.

When a project's packages have requirements that conflict with one another, it creates a situation defined as "dependency hell". Dependency hell is a common problem found in software that are built using an add-on software package or that rely on one for complete functionality. Dependency hell can take many forms and occur for many reasons, such as the need to install add-on software libraries, the need for long chains of installations, problems with a conflicting program, the creation of circular dependencies and more.
Often, rather than "reinventing the wheel", software should be designed to take advantage of other software components that are already available, or have already been designed and implemented for use elsewhere.

Definition of package engine is based on next set of operations that executes the commands and macros in the preparation section of the manifest file, checks the contents of the manifest, executes the commands and macros.

On current stage of research, it is really important  to define minimal requirements, definition and format of a component, package and manifest file. Our goal is to proceed ahead with a problem of modularization and separability of ROOT core part and other various functionality of ROOT. It gives to users more flexibility and possibility to combine various feature builds without rebuilding whole ROOT from beginning. Separation of code on components and packages could allow to have a way easily to associate functionality with a component you would like to use in ROOT.

Minimal requirement for ROOT package manager is to be able to define the dependency and its version through the manifest of package. These could be done either for ROOT packages or for external dependencies (compression algorithms packages for example). This should be a basic schema for distribution packages that are stored in  https://github.com/ and defined as dependencies.

Maximum requirements for package manager is a next set of features (the list is inspired by functionality of Swift Package Manager \cite{swift}, which is supported by a large Swift community):
\begin{enumerate}
\item Automated testing; 
\item Support of cross-platform packages;
\item Support for other build systems  and operation system package managers(homebrew and etc.);
\item Support for version control systems;
\item Standardized licensing;
\item Introduction of a package index;
\item Importing dependencies by source URL;
\item Component inter-dependency determination;
\item Complex dependency resolution.
\end{enumerate}

Important part of design is possibility to use Clang C++ modules technology. Clang C++ modules are the precompiled headers that can optimize header parsing, providing loading on-demand code from C++ modules. It is similar to ROOT precompiled header (PCH), but in the same time C++ modules are separable and modular by design. ROOT runtime C++ modules can solve a limitation ROOT PCH can be only one in the system. In the same time while using C++ modules for package management, we work on the solution of a global problem of distribution C++ modules.

Reusing concepts from introduction C++ module infrastructure for ROOT, we follow the ideas of Swift Package Manager \cite{swiftpm} and implement a standalone tool for ROOT modularization. Runtime C++ modules are available in ROOT 6.16 release as a technological preview.

\section{root-get prototype}

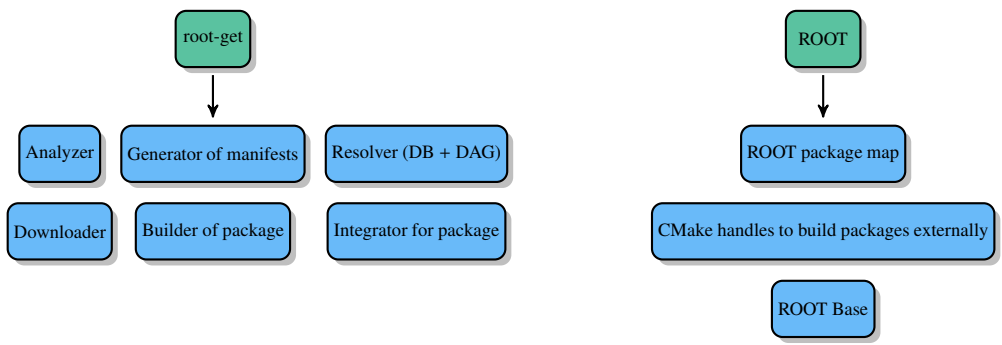
\begin{figure}[!h]
  \centering
  \begin{tikzpicture}[outer sep=0.05cm, node distance=0.8cm, scale=0.7, transform shape]
        
    \node[model, fill=my_green, name=rootget] (rootget) {root-get};
    \node[model, fill=my_green, name=root, right=10cm of rootget] (root) {ROOT};
    \node[model, fill=my_lightblue, name=gen, below=1cm of rootget] (gen) {Generator of manifests};
    \node[model, fill=my_lightblue, name=anal, left=0.3cm of gen] (anal) {Analyzer};
    \node[model, fill=my_lightblue, name=resolv, right=0.3cm of gen] (resolv) {Resolver (DB + DAG)};
    \node[model, fill=my_lightblue, name=downl, below=0.3cm of anal] (downl) {Downloader};
    \node[model, fill=my_lightblue, name=build, below=0.3cm of gen] (build) {Builder of package};
    \node[model, fill=my_lightblue, name=int, below=0.3cm of resolv] (int) {Integrator for package};
    \node[model, fill=my_lightblue, name=rpm, below=1cm of root] (rpm) {ROOT package map};
    \node[model, fill=my_lightblue, name=cmake, below=0.3cm of rpm] (cmake) {CMake handles to build packages externally};
    \node[model, fill=my_lightblue, name=rb, below=0.3cm of cmake] (rb) {ROOT Base};

    \draw[line, ->] (rootget.south) -- (gen);
    \draw[line, ->] (root.south) -- (rpm);

  \end{tikzpicture}
  \caption{Components of root-get prototype.}
  \label{fig:InformationFlow}
\end{figure}

We provide a tool that help to provide dependency management system for ROOT (Figure 2). It consists of multiple modules that provide various functionality for package management:
\begin{enumerate}
\item Analyzer - defines environment variables, checking if we have already existing manifest(package).yml files and preparing for generation routine: discovery of path for modules and packages, preparation for manifest’s generation.
\item Generator - CMake routine  in ROOT for recording info for manifest files
We are able to configure ROOT modules and packages outside of ROOT using  pair of CMake files containing all information about ROOT macro and ROOT external dependencies. In this case we are using CMake to generate ROOT package manifests for ROOT

\item Downloader - routines helping to download packages from Github or other location.
\item Resolver - provides a package management database generation module and resolution of dependencies via direct acyclic graph (DAG).
\item Builder - provides ROOT packaging scripts.
\item Intergrator - installation and deployment routine for ROOT package. 
\end{enumerate}


\section{Conclusions}

In these article we described package management ecosystem for ROOT. We defined a minimal and full set of requirements requirement for ROOT package manager. All ideas was adopted in a preliminary prototype that can download and install packages. A prototype can be connected to ROOT runtime and serve as a runtime dependency management tool.

\section{Acknowledgment}

This work has been supported by U.S. National Science Foundation grants OAC-1450377, OAC-1450323 and PHY-1624356.


\begin{thebibliography}{plain}

\bibitem{root}
Rene Brun and Fons Rademakers, ROOT - An Object Oriented Data Analysis Framework, Proceedings AIHENP'96 Workshop, Lausanne, Sep. 1996, Nucl. Inst. and Meth. in Phys. Res. A 389 (1997) 81-86.

\bibitem{rootcxxmodules}
Vassilev, Vassil. (2016). Optimizing ROOT's Performance Using C++ Modules. Journal of Physics: Conference Series. 898. . 10.1088/1742-6596/898/7/072023.

\bibitem{rootgit}
GitHub. 2018. GitHub - root-project/root: The official repository for ROOT: analyzing, storing and visualizing big data. Available at: https://github.com/root-project/root. [Accessed 03 December 2018].

\bibitem{cxxmodules}
Modules — Clang 8 documentation. 2018. Modules — Clang 8 documentation. Available at: https://clang.llvm.org/docs/Modules.html. [Accessed 30 November 2018].

\bibitem{swift}
GitHub. 2018. GitHub - apple/swift: The Swift Programming Language. Available at: https://github.com/apple/swift. [Accessed 30 November 2018].

\bibitem{swiftpm}
GitHub. 2018. GitHub - apple/swift-package-manager: The Package Manager for the Swift Programming Language. [ONLINE] Available at: https://github.com/apple/swift-package-manager. [Accessed 30 November 2018].

\bibitem{pdm}
Medium. 2018. So you want to write a package manager – Sam Boyer – Medium. Available at: https://medium.com/@sdboyer/so-you-want-to-write-a-package-manager-4ae9c17d9527. [Accessed 30 November 2018].


\end{thebibliography}
\end{document}